\documentclass[twocolumn]{jpsj2} 

\newcommand{\bvec}[1]{\mbox{\boldmath{$#1$}}}

\title{Two-band Fluctuation Exchange Study on the Superconductivity of
$\beta'$-(BEDT-TTF)$_2$ICl$_2$ under High Pressure.}

\author{Tsuguhito \textsc{Nakano}\thanks{E-mail
address:tnakano@vivace.e-one.uec.ac.jp} and Kazuhiko \textsc{Kuroki}}

\inst{Department of Applied Physics and Chemistry, The University
of Electro-Communications, Chofu, Tokyo 182-8585, Japan}

\abst{We study the pressure dependence of the superconducting transition temperature of an organic superconductor 
$\beta'$-(BEDT-TTF)$_2$ICl$_2$ by applying the fluctuation exchange method 
to the Hubbard model on the original two-band lattice at $3/4$-filling 
rather than the single band model in the strong dimerization limit. 
Our study is motivated by the fact that hopping parameters evaluated
from a first-principles study suggest that the dimerization of the
BEDT-TTF molecules is not so strong especially at high pressure.
Solving the linearized Eliashberg's equation, a d$_{xy}$-wave-like
superconducting state with realistic values of $T_c$ is 
obtained in a pressure regime somewhat higher than the actual experimental 
result. These results are similar to those obtained within the single band 
model in the previous study by Kino {\it et al}. We conclude that
the resemblance to the dimer limit is due to a combination of a good Fermi
surface nesting, a large density of states near the Fermi level, and a moderate dimerization, which cooperatively enhance electron correlation
effects and also the superconducting $T_c$.}

\kword{superconductivity,$\beta'$-(BEDT-TTF)$_2$ICl$_2$,pressure
effects,multi band,FLEX}

\begin{document}
\maketitle

\section{Introduction} 

Over the past few decades, a considerable amount of studies have been
performed on organic conductors both experimentally and theoretically.
Especially, the occurrence of 
unconventional superconductivity in these materials,
where low dimensionality and/or strong electron correlation may be 
playing an important role, 
has become one of the fascinating issues in 
condensed matter physics.\cite{OSC}

\begin{figure}
\begin{center}
 \includegraphics[scale=0.4]{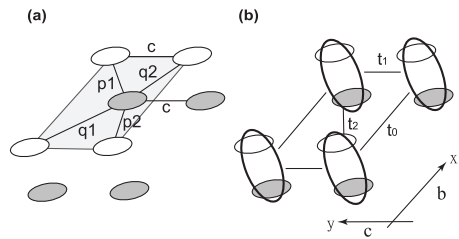}
\caption{Schematic illustration of the ET molecule layer. (a) 
The original two
 band lattice. A small oval represents an ET molecule. $p1$, $p2$,
 $\cdots$ stand for the hopping integrals $t(p1)$, $t(p2)$, $\cdots$. 
 The shaded portion denotes the unit cell. (b) The effective single band
 lattice in the dimer limit 
with effective hoppings $t_0$, $t_1$, and $t_2$. \label{fig1}}
\end{center}
\end{figure}

The title material of this paper, $\beta'$-(BEDT-TTF)$_2$ICl$_2$ 
is a charge transfer 
organic material which consists of cation BEDT-TTF (abbreviated as ET) 
molecule layers and anion ICl$_2$
layers. 
This material is a paramagnetic insulator at room temperature and ambient
pressure, and becomes an antiferromagnetic (AF) insulator below
the N\'{e}el temperature $T_N=22$ K.
Regarding the electronic structure, since two ET molecules are packed in
a unit cell as shown in Fig. \ref{fig1}(a) with 0.5 holes per ET
molecule, it is a $3/4$-filled two-band system. 
Moreover, in the $\beta'$-type arrangement of the ET molecules, which is
rather modified from the $\beta$-type because of the small size of the
anion ICl$_2$, ET molecules form dimers in the $p1$ direction, which opens up 
a gap between the two bands. Thus, only the anti-bonding band intersects
the Fermi level, so that it may be possible to look at the 
the system as a half-filled single band system. 
At ambient pressure, this picture is supported by the fact that 
system becomes an insulator despite the band being 3/4-filled.

Recently, superconductivity has been found in this material under
high pressure (above 8.2 GPa) by Taniguchi {\it et al.} It has the 
highest transition temperature $T_c$ (=$T_c^{\rm max}=14.2$ K at
$p=8.2$ GPa ) among all the molecular charge-transfer
salts. \cite{Taniguchi} Since the superconducting phase seems to sit 
next to the antiferromagnetic insulating phase in the pressure-temperature phase diagram, there is a possibility that the pairing is 
due to AF spin fluctuations. 
In fact, Kino {\it et al.} have
calculated $T_c$ and the 
N\'{e}el temperature $T_N$ using the fluctuation exchange (FLEX)\cite{Bickers} method on an effective single-band Hubbard model at
$1/2$-filling, namely the 'dimer model' obtained in the strong dimerization
limit(Fig. \ref{fig1}(b)).\cite{KKM,Kontani} In their study, the hopping
parameters of the original two band lattice are determined by fitting the
tight binding dispersion to those obtained from first principles calculation\cite{Miyazaki}, and the hopping parameters of the effective
one band lattice are obtained from a certain transformation. The obtained
phase diagram is qualitatively similar to the experimental one although
the superconducting phase appears in a higher pressure regime.

Nevertheless, we believe that it is necessary to revisit this 
problem using {\it the original two-band lattice} due to the following 
reasons.
(i)If we look into the values of the hopping integrals of the original
two-band lattice, the dimerization is not so strong, and in fact the 
gap between the bonding and the antibonding bands is only $10$ percent
of the total band width at $8$ GPa.
(ii) The effective on-site repulsion in the dimer model 
is a function of hopping integral and thus should also be a 
function of pressure. \cite{KinoFukuyama} (iii) It has been known that 
assuming the dimerization limit can result in
crucial problems as seen in the studies of $\kappa$-(ET)$_2$X, in which
the dimer model gives  $d_{x^2-y^2}$-wave pairing 
with a moderate $T_c$, while in the original four-band lattice, 
$d_{x^2-y^2}$ and $d_{xy}$-wave pairings are nearly degenerate with, 
if any, a very low $T_c$. \cite{KTAKM,KondoMoriya}
Note that in the case of $\kappa$-(BEDT-TTF)$_2$X, 
the band gap between the bonding and the antibonding band 
is more than 20 \% of the total band width,\cite{Komatsu}
which is larger than that in the title compound.

In the present paper, we calculate $T_c$ and the gap function by applying
the two-band version of FLEX for the Hubbard model on the 
two-band lattice at $3/4$-filling using the hopping parameters
determined by Miyazaki {\it et al.}  We obtain finite values of $T_c$ 
in a pressure 
regime similar to those in the single band approach. 
The present situation is in sharp contrast with the case of $\kappa$-(ET)$_2$X
 in that moderate values of $T_c$ (or we should say ``high $T_c$'' in the 
sense mentioned in \S \ref{secD4}) are obtained even in a $3/4$-filled
system, where electron correlation effects are, naively speaking, not
expected to be strong compared to true half-filled systems. 
The present study suggests that the coexistence of a good Fermi 
surface nesting,
a large density of states and a moderate (not so weak) dimerization
cooperatively enhances electron correlation effects and leads to results
similar to those in the dimer limit. 
We conclude that these factors that enhance
 correlation effects should also be the very origin of the high $T_c$ itself
of the title material.

\section{Formulation} \label{Formulation}
In the present study, we adopt a standard Hubbard Hamiltonian having 
two sites in a unit cell, where each site corresponds to an ET
molecule. The kinetic energy part of the Hamiltonian is written
as
\begin{eqnarray}
 {\cal H}_{\rm kin}&=&\sum_{i,\sigma} \biggl[ t(c)\Big( c_{(i_x,i_y+1),\sigma}^{\dagger}c_{(i_x,i_y),\sigma} \nonumber \\
&+& d_{(i_x,i_y+1),\sigma}^{\dagger}d_{(i_x,i_y),\sigma} \Big) \nonumber\\
&+& t(q1)d_{(i_x-1,i_y+1),\sigma}^{\dagger}c_{(i_x,i_y),\sigma} \nonumber\\
&+& t(q2) d_{(i_x,i_y)\sigma}^{\dagger}c_{(i_x,i_y),\sigma}\nonumber \\
&+& t(p1)d_{(i_x,i_y+1),\sigma}^{\dagger}c_{(i_x,i_y),\sigma} \nonumber \\
&+&t(p2)d_{(i_x-1,i_y),\sigma}^{\dagger}c_{(i_x,i_y),\sigma} \nonumber\\
&+& {\rm h.c} -\mu \biggr],
\end{eqnarray}
where $c_{i,\sigma}$ and $d_{i,\sigma}$ are annihilation operators of
electrons with 
spin $\sigma$ at the two different sites 
in the $i$-th unit cell, and $\mu$ represents
chemical potential. $t(p1)$, $t(p2)$, $\cdots$ are the hopping
parameters in the $p1$, $p2$, $\cdots$ directions, respectively.

The interaction part is 
\begin{eqnarray}
 {\cal H}_{\rm int} &=& U \sum_{i,\sigma}\left( n_{i,\sigma}^{c}n_{i,\sigma}^{c}+n_{i,\sigma}^d n_{i,\sigma}^d  \right), 
\end{eqnarray}
where $U$ is the on-site electron-electron interaction, and
$n_{i,\sigma}^c=c_{i,\sigma}^{\dagger}c_{i,\sigma}$ and 
$n_{i,\sigma}^d=d_{i,\sigma}^{\dagger}d_{i,\sigma}$ are the number
operators. The pressure effect on the electronic structure is introduced
through the hopping parameters within this model. As we have mentioned
above, we use the hopping parameters determined by Miyazaki {\it
et al.}\cite{Miyazaki} as shown in Table.\ref{table1}, 
which well reproduce the results
of the first principles calculation. For pressures higher than
$12$ GPa, which is the highest pressure where the first principles
calculation have been carried out, the values of the hopping integrals are
obtained by linear extrapolation as in the previous study.

In the present study, we have employed the two-band version of the fluctuation
exchange (FLEX) approximation to obtain the Green's function and the normal
self-energy. For later discussions, let us briefly review the FLEX
method, which is a kind of self-consistent random phase approximation
(RPA). Since FLEX can take large spin fluctuations into account, these
methods have been applied to the studies of high-$T_c$
cuprates and other organic superconductors.

The (renormalized) thermal Green's function
$G(\bvec{k},\varepsilon)$ is given by the Dyson's equation, 
\begin{equation}
 G^{-1}(\bvec{k},\varepsilon_n) = G_{0}^{-1}(\bvec{k},\varepsilon_n) - \Sigma(\bvec{k},\varepsilon_n), \label{Dyson}
\end{equation}
where $\varepsilon_n=(2n+1) \pi T$ is the Matsubara frequency with
$n=0, \pm 1, \pm 2, \cdots$. $G_0(\bvec{k},\varepsilon_n)$ is the
unperturbed thermal Green's function and
$\Sigma(\bvec{k},\varepsilon_n)$ is the normal self-energy, which has an
effect of suppressing $T_c$.

Using $G(\bvec{k},\varepsilon_n)$ obtained by solving eq.(\ref{Dyson}), the
irreducible susceptibility $\chi_0(\bvec{q},\omega_m)$ is given as
\begin{equation}
 \chi_0(\bvec{q},\omega_m) = 
-\frac{1}{N}\sum_{k,n}G(\bvec{k}+\bvec{q},\omega_m+\varepsilon_n)G(\bvec{k},\varepsilon_n),
\end{equation}
where $\omega_m$ is the Matsubara frequency for bosons with $m=0, \pm 1,
\pm 2, \cdots$ and 
$N$ is the number of {\boldmath$k$}-point meshes. By collecting
RPA-type diagrams, the effective interaction $V^{(1)}$ and
the singlet pairing interaction $V^{(2)}$ are obtained as 
\begin{eqnarray}
V^{(1)}(\bvec{q},\omega_m)&=& -\frac{3}{2}U^2\chi_s(\bvec{q},\omega_m)-\frac{1}{2}U^2\chi_c(\bvec{q},\omega_m)\\
V^{(2)}(\bvec{q},\omega_m)&=& U+\frac{3}{2}U^2\chi_s(\bvec{q},\omega_m)-\frac{1}{2} U^2\chi_c(\bvec{q},\omega_m),
\end{eqnarray}
where $\chi_s$,$\chi_c$ are spin and charge susceptibilities, respectively, 
given as
\begin{equation}
\chi_{s,c}(\bvec{q},\omega_m) = \frac{\chi_0(\bvec{q},\omega_m)}{1 \mp U\chi_0(\bvec{q},\omega_m)}.
\end{equation}
Then the normal self-energy is given by 
\begin{equation}
 \Sigma(\bvec{k},\varepsilon_n)=-\frac{T}{N}\sum_{q,m}G(\bvec{k}-\bvec{q},\varepsilon_n)\left[V^{(1)}(\bvec{q},\omega_m)-U^2\chi_0(\bvec{q},\omega_m)\right]. \label{selfenery}
\end{equation}
The obtained self-energy
$\Sigma(\bvec{k},\varepsilon_n)$ is fed back into the Dyson's equation
eq.(\ref{Dyson}), and by repeating these procedures, the self-consistent
$G(\bvec{k},\varepsilon_n)$ is obtained. 

\begin{table}[t]
 \begin{center}
   \caption{\small{\label{table1}}Pressure dependence of the hopping
  parameters of $\beta'$-(ET)$_2$ICl$_2$ determined by
  Miyazaki and Kino. (ref.\citen{Miyazaki})}
 \begin{tabular}[t]{|c||ccccc|} \hline 
   P &$t(p1)$&$t(p2)$&$t(q1)$&$t(q2)$&$t(c)$ \\ \hline
   0 (GPa) &-0.181 (eV)&0.0330&-0.106&-0.0481&-0.0252 \\
   4 &-0.268&0.0681&-0.155&-0.0947&-0.0291 \\
   8 &-0.306&0.0961&-0.174&-0.120&-0.0399 \\
   12&-0.313&0.142&-0.195&-0.122&-0.0347 \\ 
  16&-0.320&0.188&-0.216&-0.124&-0.0295 \\\hline
 \end{tabular}
  \end{center}
\end{table}

In the two-band version of FLEX, $G(\bvec{k},\varepsilon_n)$, $\chi_0$,
$\chi_{s,c}$, $\Sigma(\bvec{k},\omega_m)$ become $2 \times 2$ matrices,
e.g. $G_{\alpha\beta}$, where $\alpha$ and $\beta$ denote one of the two
sites in a unit cell.

Once $G_{\alpha\beta}(\bvec{k},\varepsilon_n)$ and $V_{\alpha\beta}^{(2)}$ 
are obtained by FLEX, we can calculate  $T_c$ by solving the 
linearized Eliashberg's equation as follows,
\begin{eqnarray} 
 \lambda\phi_{\alpha\beta}(\bvec{k},\varepsilon_n)=
-\frac{T}{N}\sum_{k',m,\alpha',\beta'}
V_{\alpha\beta}^{(2)}(\bvec{k}-\bvec{k}',\varepsilon_n-\varepsilon_m) \nonumber \\ \times G_{\alpha\alpha'}(\bvec{k}',\varepsilon_m)G_{\beta\beta'}(-\bvec{k}',-\varepsilon_m)\phi_{\alpha'\beta'}(\bvec{k}',\varepsilon_m), \label{Eliash}
\end{eqnarray}
where $\phi(\bvec{k})$ is the superconducting gap function. 
The transition temperature $T_c$ is
determined as the temperature where the eigenvalue $\lambda$ reaches unity.

In the actual calculation, we use $64 \times 64$ {\boldmath $k$}-point 
meshes and $16384$ Matsubara frequencies in order to ensure convergence at
the lowest temperature studied ($T/|t(p1)|=0.002$). 
The bandfilling (the number of electrons per site) is
fixed at $n=1.5$.

When $U\chi_0(\bvec{q},\omega_m=0)=1$, the spin susceptibility diverges 
and a magnetic ordering takes place.
In the FLEX calculation in two-dimensional systems, Mermin-Wagner's 
theorem is satisfied,\cite{MW,Deisz} so that $U\chi_0(\bvec{q},\omega_m=0)<1$, namely 
true magnetic ordering does not take place. However, this 
is an artifact of adopting a purely two-dimensional model, while 
the actual material is {\it quasi} two dimensional.
Thus, in the present study, we assume that 
if there were a weak three dimensionality, a  
magnetic ordering with wave vector {\boldmath$q$} would occur when 
\begin{equation}
\max_{q}\left\{U\chi_0(\bvec{q},\omega_n=0) > 0.995 \right\}, \label{AFcriterion}
\end{equation}
is satisfied in the temperature range where $\lambda <1$. 
Therefore we do not calculate $T_c$  in such a parameter regime. 
\cite{comment}

\section{Results}

\begin{figure}
 \begin{center}
 \includegraphics[scale=0.65]{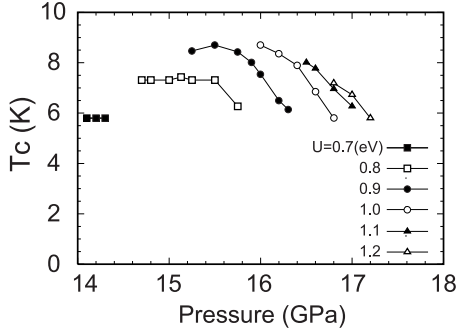}
 \caption{Transition temperature $T_c$ as functions of pressure for
  several values of $U$.\label{fig2}} 
 \end{center}
\end{figure}

Now we move on to the results. Figure \ref{fig2} 
shows the pressure dependence of $T_c$  obtained for several
values of $U$. Since our calculation is restricted to 
temperatures above $\sim$ 6 K, $T_c$ is obtained 
within that temperature range.
At pressure lower than the superconducting regime, 
the system is in the AF phase in the sense we mentioned in 
\S \ref{Formulation}. 
The maximum $T_c$ obtained is $T_c^{\rm max}=8.7$ K
(at $15.5$ GPa for $U=0.9$ eV, and $16.0$ GPa for $U=1.0$ eV), which is 
somewhat smaller than the experimental maximum value of $T_c$, but can be 
considered as fairly realistic. 
The overall phase diagram is qualitatively similar to the experimental phase diagram, but the pressure range in which superconductivity occurs 
is above $\sim 14$ GPa and extends up to $\sim 17$ GPa or higher,
is higher than the experimental results.
These results  are similar to those obtained within the dimer model
approach.\cite{KKM}


\begin{figure}[t]
\begin{center}
 \includegraphics[scale=0.4]{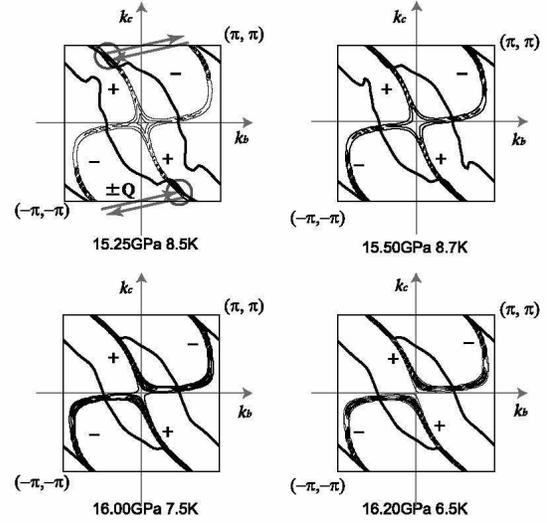}
 \caption{Contour plots of $|G(\bvec{k})|^2$ and the nodes of the 
superconducting gap function 
$\phi(\bvec{k})$ for $U=0.9$ eV at $p=15.25$, $15.5$, $16.0$, 
and $16.2$ GPa. At $p=15.25$ GPa, $\phi(\bvec{k})\phi(\bvec{k}+\bvec{Q})$
 is positive at the circled portions of the Fermi surface. \label{fig3}}
\end{center} 
\end{figure}

\begin{figure}
 \begin{center}
  \includegraphics[scale=0.4]{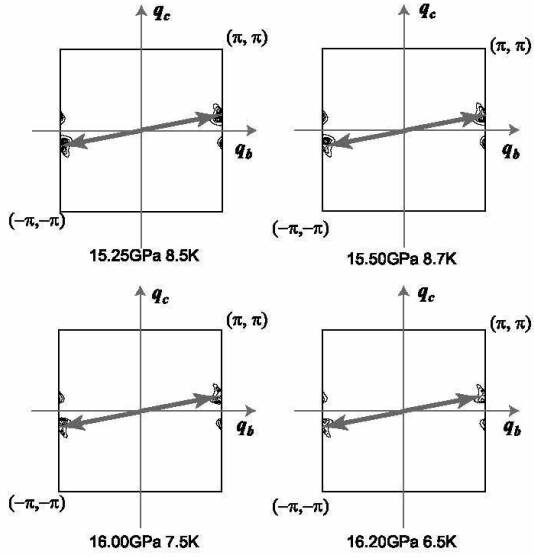}
  \caption{Contour plots of the spin susceptibility $\chi_s(\bvec{q})$ for
  the same values of $U$ and $p$ as in Fig. \ref{fig3}.\label{fig4}}
 \end{center}
\end{figure}

Figure \ref{fig3} shows the nodal lines of $\phi(\bvec{k})$ and 
the contour plots of $|G(\bvec{k})|^2$  
for several values of pressure with $U=0.9$ eV.
In the plots of $|G(\bvec{k})|^2$, the center of the 
densely bundled contour lines correspond to the ridges of $|G(\bvec{k})|^2$
and thus the Fermi surface, while the thickness of these bundles 
can be considered as a measure for the density of states near the 
Fermi level, namely, the thicker these 
bundles, the larger number of states lie near the Fermi level.
With increasing pressure, 
the Fermi surface changes its topology from a one dimensional
one open in the $\bvec{k_c}$ direction to a closed two dimensional one around 
$(0,\pi)$. Again like in the single band approach, the pairing symmetry is
$d_{xy}$-wave-like in the sense that $\phi(\bvec{k})$ changes 
its sign as ($+-+-$) along the Fermi surface and the
nodes of the gap intersect the Fermi surfaces near $x$ and $y$
axes. The peak position of the spin susceptibility $\chi_s(\bvec{q})$ shown in
Fig. \ref{fig4}, which should correspond to the nesting vector of the
Fermi surface, stays around $\bvec{Q}=(\pi,\pi/4)$ regardless of
the pressure. This vector $\bvec{Q}$ bridges the portion of the
Fermi surface with $\phi({\bvec{k}})<0$ and $\phi(\bvec{k}+\bvec{Q})>0$, which is the origin of the $d_{xy}$-wave like gap.



\section{Discussion}

In this section, we physically interpret some of our calculation results.

\subsection{Pressure dependence of $T_c$ at fixed values of $U$}

For fixed values of $U$, $T_c$ tends to be suppressed at high pressure
as seen in Fig. \ref{fig2}. This can be explained as follows.
We have seen in Fig. \ref{fig4} that the peak position of the 
spin susceptibility does not depend on  pressure, 
but the peak value itself 
decreases with pressure for fixed values of $U$ as shown in Fig. \ref{fig6}.  
This is because the nesting of the Fermi surface becomes 
degraded due to the dimensional crossover of the Fermi surface 
mentioned previously. 
Consequently, the pairing interaction $V^{(2)}$ (nearly proportional to
the spin susceptibility) becomes smaller, 
so that $T_c$ becomes lower, with increasing pressure.

\begin{figure}[h]
\begin{center}
\includegraphics[scale=0.4]{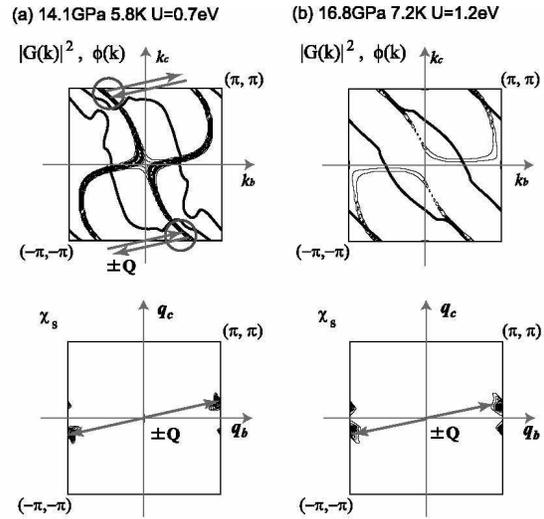}
\caption{Contour plots of $|G(\bvec{k})|^2$, the nodes of the gap 
function, and the spin susceptibility at (a)$U=0.7$ eV, $p=14.1$ GPa. 
(b)$U=1.2$ eV,$p=17.2$ GPa. 
At the circled portions of the Fermi surface, 
$\phi(\bvec{k})\phi(\bvec{k}+\bvec{Q})$ is positive.\label{fig5}}
\end{center}
\end{figure}

For $U=0.9$ eV (and possibly for $U=0.8$ eV), 
$T_c$ is slightly suppressed also at low pressure,
so that a optimum pressure exists.
This may be due to the fact that at low pressure, $\bvec{Q}$ 
(spin susceptibility peak position) bridges some portions of the Fermi 
surface that has the same sign of the gap (Fig. \ref{fig3}(a)). In fact, this tendency of $\bvec{Q}$ bridging
the same gap sign is found 
to be even stronger for lower values of pressure as seen in Fig. \ref{fig5}(a),
while at higher pressure, the nodes of the 
gap run along the Fermi surface so as to suppress this 
tendency (Fig. \ref{fig5}(b)). We will come back to this point 
in \S \ref{secD3}.

\subsection{Pressure dependence of the maximum $T_c$ upon varying $U$}

As can be seen from Fig. \ref{fig2}, for each value of pressure, 
there exists an optimum value of $U(=U_{\rm opt})$ 
which maximizes $T_c$ .
In this subsection, let us discuss the pressure dependence of this 
optimized $T_c$ as a function of $U_{\rm opt}$, that is,  $T_c(U_{\rm opt})$.
Regarding the lower pressure region, $\chi_0$ is large because of the
good nesting of the Fermi surface, so that $U$ has to be small in order
to avoid AF ordering (in the sense mentioned in \S \ref{Formulation}). 
This is the reason why the maximum value of $T_c$ is relatively low 
in the low pressure regime as seen in Fig. \ref{fig2}. On the other hand,
in the high pressure region, $\chi_0$ is small because of the 2D-like
Fermi surfaces, so that $U$ must be large in order to have large $\chi$
and thus pairing interaction. Such a large $U$, however, makes the
normal self energy $\Sigma(\bvec{k},\varepsilon_n)$ large, which again results
in low $T_c$ (In Fig. \ref{fig5}.(b),the low height of $|G(\bvec{k})|^2$
represents the large effective mass of Fermion.). Thus, relatively high
$T_c(U_{\rm opt})$ is obtained at some intermediate values of pressure.

\subsection{Pressure dependence of the gap function} \label{secD3}

In this subsection we discuss the variation of the gap function with
increasing pressure. To understand this variation in real space, we use
the following relation 
\begin{eqnarray}
 O&=&\sum_{k}\phi(\bvec{k})c_{k \uparrow}c_{-k \downarrow} \nonumber \\
&=&\sum_{i,\delta}g(\delta)(c_{i \uparrow}c_{i+\delta \downarrow}-c_{i \downarrow}c_{i+\delta \uparrow}),
\end{eqnarray}
where $i$ and $i+\delta$ denote sites in real space where pairs are formed,
and $g(\delta)$ is a weight of such pairing. Note that a `site' here  
corresponds to a unit cell (or a dimer). Considering up to 22$^{nd}$
nearest neighbor pairings, we have determined a set of $g(\delta)$ that
well reproduces $\phi(\bvec{k})$ obtained by the FLEX, using least squares
fitting, as shown typically in Fig. \ref{fig7}.

\begin{figure}[t]
\begin{center}
\includegraphics[scale=0.65]{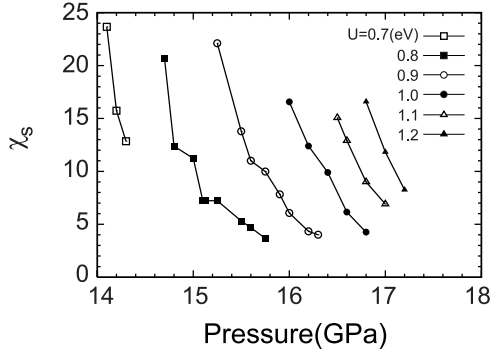}
\caption{Pressure dependence of the maximum values of $\chi_s(\bvec{q},\omega_m=0)$ 
for several values of $U$.\label{fig6}}
\end{center}
\end{figure}

The result of this analysis is shown in Fig. \ref{fig7}, where the thickness of
the lines represents the {\it weight} of the pairing determined from
the value of $g(\delta)$. We can see that the direction in which the
dominant pairings take place changes from {\boldmath $b$} to
{\boldmath $b$}$+${\boldmath $c$} as the pressure increases, 
which looks more like $d_{xy}$-wave like pairing.
These changes of the pairing directions are correlated 
with the increasing of hopping
$t(p2)$. Thus, from the viewpoint of this real space analysis, we can
say that the change of the dominant pairing direction due to the
increase of $t(p2)$ suppress the tendency of the nesting vector
{\boldmath $Q$} bridging the portions of Fermi surface with the same gap
sign.

\subsection{Origin of the ``high $T_c$''}
\label{secD4}

Finally, we discuss the reason why the obtained results, 
namely the values of $T_c$ and the form of the gap function,  
resemble that of the dimer limit approach. 
The present situation is in sharp contrast with the case of 
$\kappa$-(BEDT-TTF)$_2$X, where it has been known that, 
compared to the results of the dimer limit approach, 
\cite{KondoMoriya2,Schmalian,KinoKontani}
the position of the gap nodes changes and the values of $T_c$, if any,
is drastically reduced in the original four band model with moderate 
dimerization\cite{KondoMoriya,KTAKM}.

A large difference between the present case and $\kappa$-(BEDT-TTF)$_2$X
is the Fermi surface nesting. As mentioned previously, 
the quasi-one-dimensionality of the system gives good Fermi surface 
nesting with strong spin fluctuations, 
fixing the nesting vector and thus the pairing symmetry firmly, 
while in the case of  $\kappa$-(BEDT-TTF)$_2$X, the Fermi surface has 
no good Fermi surface nesting. 

\begin{figure}[t]
 \begin{center}
  \includegraphics[scale=0.55]{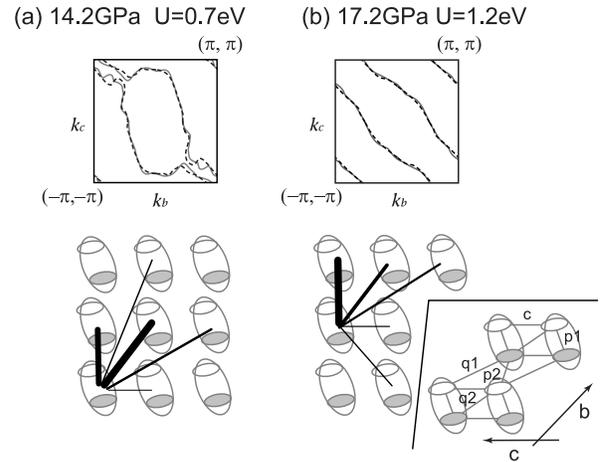}
  \caption{(upper panel) Results of the least squares fitting of the gap
  function. The dashed curve is the result of the FLEX calculation and the
  solid curve is the fitting. (lower panel) The weight of the directions
  in which the pairings take place for 
  (a)$p=14.2$ GPa, $U=0.7$ eV (b) $p=17.2$ GPa, $U=1.2$ eV\label{fig7}}
 \end{center}
\end{figure}

\begin{figure}
 \begin{center}
  \includegraphics[scale=0.55]{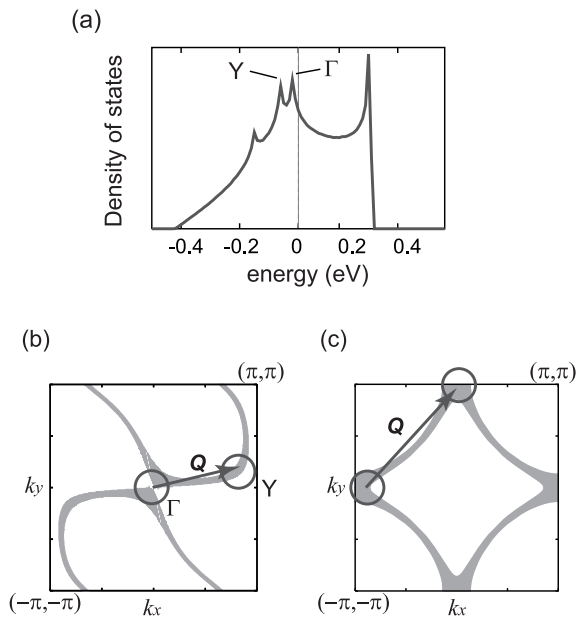}
  \caption{(a) The density of states of the antibonding band 
at $p=17$ GPa for $U=0$. 
(b) The Fermi surface with finite thickness defined by 
$E_F-\delta E<E(k_x,k_y)<E_F+\delta E_F$, where $E(k_x,k_y)$ is the
  band dispersion, and $\delta E_F=0.015$ eV is taken here. (c) A similar
  plot for the cuprates.\label{fig8}}
 \end{center}
\end{figure}

However, it is unlikely that the good Fermi surface 
nesting alone can account for the resemblance between the 
3/4-filled model and the dimer limit model 
because, for example, in the study for another organic superconductor  
(TMTSF)$_2$X, it has been known 
that a 1/4-filled model with no dimerization 
gives weak spin fluctuations within FLEX 
even though the Fermi surface nesting is very good.\cite{KinoKontani2}
One difference from the case of (TMTSF)$_2X$ is the 
presence of moderate (not so weak) dimerization, but there is also a 
peculiar structure in the density of states as pointed out in the 
previous study.\cite{KKM}
Fig. \ref{fig8}(a) shows the density of states of the antibonding band 
at $p=17$ GPa for $U=0$. 
The two peaks near the Fermi level (energy=0) originates 
from the saddle points of the band dispersion located 
at the $\Gamma$ point and the Y point ($\bvec{k}=(\pi,0)$).
Consequently, the ``Fermi surface with finite thickness'',
defined by $E_F-\delta E<E(k_x,k_y)<E_F+\delta E$,  
becomes thick near the $\Gamma$ and the Y points, as shown 
in Fig. \ref{fig8}(b). In fact, this trend is already 
seen in the contour plots of the 
Green's function in Figs. \ref{fig3} and \ref{fig5}, 
where the bundles of the contour lines become thick near the $\Gamma$ and/or 
the Y points.
Importantly, the wave vector (the nesting vector $\simeq (\pi,\pi/4)$) at which the 
spin fluctuations strongly develop bridges the states near the $\Gamma$ point and those somewhat close to 
the Y point (Fig. \ref{fig8}(b)), so that there are many states
which contribute to the pair scattering. 
From the above argument, 
our results suggest that the {\it coexistence} of 
the good Fermi surface nesting, the large density of states near the Fermi level, 
and the moderate dimerization cooperatively enhances electron
correlation effects, thereby giving results similar to those in the
dimer (strong correlation) limit.

Now, these factors that enhance electron correlation should 
also make $T_c$ itself rather high. 
In fact, $T_c$ of $\sim 0.0006W$ almost reached in the present model, 
where $W$ is the band width (around $1.3$ eV for $p=16$ GPa), 
is relatively high among $T_c$ obtained by FLEX+Eliashberg equation 
approach in 
various Hubbard-type models. Namely, Arita {\it et al.} have
previously shown\cite{AKA} that $T_c$ of order $0.001W$ is about the
highest we can reach within the Hubbard model\cite{exception}, which is
realized on a two dimensional square lattice near half filling, namely,
a model for the high $T_c$ cuprates. 
The present study is in fact reminiscent of the FLEX study of the 
high $T_c$ cuprates, where the 3/4-filled two
band model\cite{Koikegami} and the half-filled single band model indeed
give similar results on the superconducting $T_c$ and the pairing
symmetry.\cite{Bickers} The cuprates also have a large density of 
states at the Fermi level originating from the 
so called hot spots around $(\pi,0)$ and $(0,\pi)$, 
and the wave vector $\sim (\pi,\pi)$ at which 
the spin fluctuations develop bridges these hot spots, as 
shown in Fig. \ref{fig8}(c).
Moreover, a moderate 
band gap also exists in the cuprates 
between the fully filled bonding/non-bonding bands and the
nearly half-filled antibonding band. The situation is thus 
somewhat similar to the present case.
To conclude this section, it is highly likely
that the coexistence of the factors that enhance correlation
effects and thus make the results between the 3/4-filled original model
and the half-filled dimer model similar is the very reason for the
``high $T_c$'' in the title material.

\section{Conclusion}
In the present paper, we have studied the pressure dependence of the 
superconducting transition temperature of an organic superconductor
$\beta'$-(BEDT-TTF)$_2$ICl$_2$ by applying two-band version of FLEX 
to the original two-band Hubbard model at $3/4$-filling with the 
hopping parameters determined from first principles calculation. 
The good Fermi surface nesting, 
the large density of states, and the moderate dimerization 
cooperatively enhance electron correlation effects, thereby 
leading to results similar to those in the dimer limit.
We conclude that these factors that enhance electron correlation is 
the origin of the high $T_c$ in the title material.

As for the discrepancy between the present result and the experiment
concerning the pressure regime where the superconducting phase appears,
one reason may be due to the fact that we obtain $T_c$ only when $U\chi_0 >
0.995$ is not satisfied despite the fact that 
this criterion for ``antiferromagnetic
ordering'', originally adopted in the dimer model approach,\cite{KKM}
does not have a strict quantitative basis. Therefore, it may be possible
to adopt, for example, $U\chi_0 > 0.999$ as a criterion for
``antiferromagnetic ordering'', which will extend the superconducting
phase into the lower pressure regime. Nevertheless, it seems that such
consideration alone cannot account for the discrepancy because in order
to ``wipe out'' the superconductivity in the high pressure regime as in
the experiments, smaller values of $U$ would be necessary, which would
give unrealistically low values of $T_c$. Another possibility for the
origin of the discrepancy may be due to the ambiguity in determining the
hopping integrals from first principles calculation. Further qualitative
discussion may be necessary on this point in the future study.


\section*{Acknowledgment }
We are grateful to Ryotaro Arita for various discussions.
The numerical calculation has been done at the Computer Center,
ISSP, University of Tokyo. This study has been supported by
Grants-in-Aid for Scientific Research from the Ministry of Education,
Culture, Sports, Science and Technology of Japan, and from the Japan
Society for the Promotion of Science.


\begin {thebibliography}{99}
\bibitem{comment}
In previous studies, the N\'{e}el temperature $T_N$ is determined by a  
condition similar to  eq.(\ref{AFcriterion}). 
In the present study, we do not evaluate $T_N$ because 
the antiferromagnetic phase in the actual material occurs 
below the Mott transition temperature (or the antiferromagnetic phase is 
within the Mott insulating phase), while a Mott transition cannot be 
treated within the present approach.
\end{thebibliography}
\end{document}